\documentstyle[11pt]{article}
\begin{document}

\title{{\bf Holographic dark energy in a non-flat universe with Granda-Oliveros cut-off}}
\author{K. Karami$^{1,2}$,\thanks{E-mail: KKarami@uok.ac.ir}\\
J. Fehri${^1}$\\$^{1}$\small{Department of Physics, University of
Kurdistan, Pasdaran St., Sanandaj, Iran}\\$^{2}$\small{Research
Institute for Astronomy
$\&$ Astrophysics of Maragha (RIAAM), Maragha, Iran}\\
}

\maketitle

\begin{abstract}
Motivated by Granda and Oliveros (GO) model, we generalize their
work to the non-flat case. We obtain the evolution of the dark
energy density, the deceleration and the equation of state
parameters for the holographic dark energy model in a non-flat
universe with GO cut-off. In the limiting case of a flat universe,
i.e. $k = 0$, all results given in GO model are obtained.
\end{abstract}
\clearpage
\section{Introduction}
Type Ia supernovae observational data suggest that our universe is
experiencing an accelerated expansion driven by an exotic energy
with negative pressure which is so-called dark energy (DE)
\cite{Riess}. However, the nature of DE is still unknown, and
people
 have proposed some candidates to describe it. The cosmological
 constant, $\Lambda$, is the most obvious theoretical candidate of
 DE, which has the equation of state $\omega=-1$.
 Astronomical observations indicate that the cosmological constant is many orders of magnitude
 smaller than estimated in modern theories of elementary particles \cite{Weinberg}. Also the
 "fine-tuning" and the "cosmic coincidence" problems are the two
 well-known difficulties of the cosmological constant problems
 \cite{Copeland}.

 There are different alternative theories for the dynamical DE scenario which have been
 proposed by people to interpret the accelerating universe. i) The scalar field models of DE including
 quintessence \cite{Wetterich},
 phantom (ghost) field \cite{Caldwell1}, K-essence \cite{Chiba}
 based on earlier work of K-inflation \cite{Picon3},
 tachyon field \cite{Sen}, dilatonic ghost condensate \cite{Gasperini},
 quintom \cite{Elizalde1}, and so forth. ii) The DE models including
 Chaplygin gas \cite{Kamenshchik}, braneworld models \cite{Deffayet}, agegraphic DE models \cite{Cai},
 and $f(R)$-gravity models \cite{Capozziello}, etc.

Recently, a new DE candidate, based on the holographic principle,
was proposed \cite{Horava}. According to the holographic
principle, the number of degrees of freedom in a bounded system
should be finite and has relations with the area of its boundary
\cite{Hooft}. In quantum field theory a short distance (UV)
cut-off $\Lambda$ is related to a long distance (IR) cut-off $L$
due to the limit set by formation of a black hole, which results
in an upper bound on the zero-point energy density \cite{Cohen}.
By applying the holographic principle to cosmology, one can obtain
the upper bound of the entropy contained in the universe
\cite{Fischler}. Following this line, Li \cite{Li} argued that for
a system with size $L$ and UV cut-off $\Lambda$, it is required
that the total energy in a region of size $L$ should not exceed
the mass of a black hole of the same size, thus
$L^3\rho_{\Lambda}\leq LM_P^2$, where $\rho_{\Lambda}$ is the
quantum zero-point energy density caused by UV cut-off $\Lambda$
and $M_P$ is the reduced Planck Mass $M_P^{-2}=8\pi G$. The
largest $L$ allowed is the one saturating this inequality, thus
$\rho_{\Lambda}=3c^2M_P^2L^{-2}$, where $c$ is a numerical
constant. Recent observational data, which have been used to
constrain the holographic DE model, show that for the non-flat
universe $c=0.815_{-0.139}^{+0.179}$ \cite{Li5}, and for the flat
case $c=0.818_{-0.097}^{+0.113}$ \cite {Li6}. Also Li \cite{Li}
showed that the cosmic coincidence problem can be resolved by
inflation in the holographic DE model, provided the minimal number
of e-foldings \cite{Li}. The holographic models of DE have been
studied widely in the literature
\cite{Enqvist,Elizalde2,Guberina}. As we mentioned before, the UV
cut-off is related to the vacuum energy, and IR cut-off is related
to the large scale of the universe, for example Hubble horizon,
future event horizon or particle horizon. Taking $L$ as the size
of the current universe, for instance, the Hubble scale, the
resulting energy density is comparable to the present day DE.
However, as found by Hsu \cite{Hsu}, in that case, the evolution
of the DE is the same as that of dark matter (dust matter), and
therefore it cannot drive the universe to accelerated expansion.
The same appears if one chooses the particle horizon of the
universe as the length scale $L$ \cite{Li}. An interesting
proposal is made by Li \cite{Li}: Choosing the event horizon of
the universe as the length scale, the holographic DE not only
gives the observation value of DE in the universe, but also can
drive the universe to an accelerated expansion phase. In that
case, however, an obvious drawback concerning causality appears in
this proposal. Event horizon is a global concept of space–time;
existence of event horizon of the universe depends on future
evolution of the universe; and event horizon exists only for
universe with forever accelerated expansion. This motivated GO
\cite {Granda1} to propose a new infrared cut-off for the
holographic DE, which besides the square of the Hubble scale also
contains the time derivative of the Hubble scale. This model
depends on local quantities and avoids the problem of causality
which appears using the event horizon area as the IR cut-off. They
obtained the evolution of both the deceleration and the equation
of state parameters for this model in a flat universe.

Besides, as usually believed, an early inflation era leads to a
flat universe. This is not a necessary consequence if the number
of e-foldings is not very large \cite{Huang}. It is still possible
that there is a contribution to the Friedmann equation from the
spatial curvature when studying the late universe, though much
smaller than other energy components according to observations.
Therefore, it is not just of academic interest to study a universe
with a spatial curvature marginally allowed by the inflation model
as well as observations. Some experimental data have implied that
our universe is not a perfectly flat universe and that it
possesses a small positive curvature \cite{Bennett}. The tendency
of preferring a closed universe appeared in a suite of CMB
experiments \cite{Sievers}. The improved precision from WMAP
provides further confidence, showing that a closed universe with
positively curved space is marginally preferred \cite{Uzan}. In
addition to CMB, recently the spatial geometry of the universe was
probed by supernova measurements of the cubic correction to the
luminosity distance \cite{Caldwell}, where a closed universe is
also marginally favored.

All mentioned in above motivate us to consider the holographic DE
model with the new infrared cut-off proposed by \cite{Granda1} and
extend their work to a non-flat case. To do this, in Section 2, we
obtain the evolution of the DE density, the deceleration and the
equation of state parameters for the holographic DE model given by
\cite{Granda1} in the context of the non-flat universe. In Section
3, we give numerical results. Section 4 is devoted to conclusions.

\section{Holographic DE in non-flat FRW universe with GO cut-off}
Following GO \cite{Granda1}, the holographic DE density is given
by
\begin{equation}
\rho_{\Lambda}=3(\alpha H^{2}+\beta\dot{H}),\label{rholambda1}
\end{equation}
where $\alpha$ and $\beta$ are constants. GO \cite{Granda1} argued
that since the underlying origin of the holographic DE is still
unknown, the inclusion of the time derivative of the Hubble
parameter may be expected as this term appears in the curvature
scalar, and has the correct dimension. This kind of density may
appear as the simplest case of more general $f(H,\dot{H})$
holographic density in the FRW background. Comparing Eq.
(\ref{rholambda1}) with the holographic DE density
$\rho_{\Lambda}=3c^2M_P^2L^{-2}$ shows that the corresponding IR
cut-off $L$ for the model (\ref{rholambda1}) is
\begin{equation}
L=H^{-1}\Big(1+\frac{\beta}{\alpha}\frac{\dot{H}}{H^2}\Big)^{-1/2},\label{L}
\end{equation}
which depends on local quantities and avoids the problem of
causality which appears using the event horizon area as the IR
cut-off. For the special case $\beta=0$, Eq. (\ref{L}) yields the
Hubble horizon as the IR cut-off, i.e. $L=H^{-1}$.

The first Friedmann equation in non-flat universe is given by
\begin{equation}
H^{2}=\frac{1}{3}(\rho_{\rm m}+\rho_{\rm r}+\rho_{\Lambda}+\rho_{\rm
k}),\label{F1}
\end{equation}
where we take $8\pi G=1$ and $\rho_{\rm k}=-3k/a^2$. Parameter $k$
denotes the curvature of space $k=0,1,-1$ for a flat, closed and
open universe, respectively. Let us define the current densities
parameters for $a_0=1$ as usual
\begin{equation}
\Omega_{\rm m_0}=\frac{\rho_{\rm m_0}}{3H_{0}^{2}},~~~\Omega_{\rm
r_0}=\frac{\rho_{\rm
r_0}}{3H_{0}^{2}},~~~\Omega_{\Lambda_0}=\frac{\rho_{\Lambda_0}}{3H_{0}^{2}},~~~\Omega_{\rm
k_0}=\frac{\rho_{\rm
k_0}}{3H_{0}^{2}}=-\frac{k}{H_{0}^{2}}\label{Omegas},
\end{equation}
where these densities satisfy $\Omega_{\rm m_0}+\Omega_{\rm
r_0}+\Omega_{\Lambda_0}+\Omega_{\rm k_0}=1$. A closed universe
with a small positive curvature ($\Omega_{\rm k_0}\sim -0.015$) is
compatible with observations \cite{Bennett}.

Now we can rewrite the first Friedmann equation in terms of $x=\ln
a$ as
\begin{equation}
\tilde{H}^{2}=\Omega_{\rm m_0}e^{-3x}+\Omega_{\rm
r_0}e^{-4x}+\Omega_{\rm k_0}e^{-2x}+\alpha
\tilde{H}^{2}+\frac{\beta}{2}\frac{d \tilde{H}^{2}}{d x},\label{F2}
\end{equation}
where $\tilde{H}=\frac{H}{H_{0}}$ is the scaled Hubble expansion
rate and $H_{0}$ is the present value of the Hubble parameter (for
$x=0$).

Solving Eq. (\ref{F2}), we obtain
\begin{equation}
\tilde{H}^{2}=\frac{2}{3\beta-2\alpha+2}\Omega_{\rm m_0}e^{-3x}
+\frac{1}{2\beta-\alpha+1}\Omega_{\rm r_0}e^{-4x}+
\frac{1}{\beta-\alpha+1}\Omega_{\rm
k_0}e^{-2x}+Ce^{-2x(\alpha-1)/\beta},\label{F3}
\end{equation}
where $C$ is an integration constant. From Eqs. (\ref{rholambda1})
and (\ref{F3}), the scaled holographic DE density,
$\tilde{\rho}_{\Lambda}=\frac{\rho_{\Lambda}}{3H_{0}^{2}}$, can be
obtained as
\begin{equation}
\tilde{\rho}_{\Lambda}=\frac{3\beta-2\alpha}{2\alpha-3\beta-2}\Omega_{\rm
m_0}e^{-3x}+\frac{2\beta-\alpha}{\alpha-2\beta-1}\Omega_{\rm
r_0}e^{-4x}+\frac{\beta-\alpha}{\alpha-\beta-1}\Omega_{\rm
k_0}e^{-2x} +Ce^{-2x(\alpha-1)/\beta}.\label{rholambda2}
\end{equation}
Using Eqs. (\ref{F3}) and (\ref{rholambda2}), the DE density
parameter,
$\Omega_{\Lambda}=\rho_{\Lambda}/3H^2=\tilde{\rho}_{\Lambda}/\tilde{H}^{2}$,
can be obtained in terms of redshift $z=\frac{1}{a}-1$ as
\begin{equation}
\Omega_{\Lambda}=\frac{\frac{3\beta-2\alpha}{2\alpha-3\beta-2}\Omega_{\rm
m_0}(1+z)^3+ \frac{2\beta-\alpha}{\alpha-2\beta-1}\Omega_{\rm
r_0}(1+z)^4+\frac{\beta-\alpha}{\alpha-\beta-1}\Omega_{\rm
k_0}(1+z)^2
+C(1+z)^{\frac{2(\alpha-1)}{\beta}}}{\frac{2}{3\beta-2\alpha+2}\Omega_{\rm
m_0}(1+z)^3+\frac{1}{2\beta-\alpha+1}\Omega_{\rm r_0}(1+z)^4
+\frac{1}{\beta-\alpha+1}\Omega_{\rm
k_0}(1+z)^2+C(1+z)^{\frac{2(\alpha-1)}{\beta}}}.\label{OmegaLambda}
\end{equation}
Using Eq. (\ref{rholambda2}) and the holographic DE conservation
equation
\begin{equation}
\tilde{p}_{\Lambda}=-\tilde{\rho}_{\Lambda}-\frac{1}{3}\frac{d
\tilde{\rho}_{\Lambda}}{d x},
\end{equation}
we obtain the DE pressure
\begin{eqnarray}
\tilde{p}_{\Lambda}=\frac{2\alpha-3\beta-2}{3\beta}Ce^{-2x(\alpha-1)/\beta}
+\frac{2\beta-\alpha}{3(\alpha-2\beta-1)}\Omega_{\rm
r_0}e^{-4x}+\frac{\beta-\alpha}{3(\beta-\alpha+1)}\Omega_{\rm
k_0}e^{-2x}.\label{plambda}
\end{eqnarray}
Note that for the special case $\beta=0$, using Eqs.
(\ref{rholambda2}) and (\ref{plambda}), the equation of state
(EoS) parameter of the DE,
$\omega_{\Lambda}=\tilde{p}_{\Lambda}/\tilde{\rho}_{\Lambda}$, is
obtained as
\begin{equation}
\omega_{\Lambda}=\frac{\Omega_{\rm r_0}(1+z)^4-\Omega_{\rm
k_0}(1+z)^2}{3[\Omega_{\rm m_0}(1+z)^3+\Omega_{\rm
r_0}(1+z)^4+\Omega_{\rm k_0}(1+z)^2]},\label{wbeta0}
\end{equation}
which shows that neglecting the contributions from radiation and
curvature, i.e. $\Omega_{\rm r_0}=\Omega_{\rm k_0}=0$, yields the
pressureless DE, i.e. $\omega_{\Lambda}=0$, where its EoS behaves
like the (dark matter) dust matter. This result is same as that
obtained by Hsu \cite{Hsu} for the holographic DE model with the
IR cut-off $L=H^{-1}$. Therefore choosing the Hubble horizon as
the IR cut-off cannot drive the universe to accelerated expansion.

Using $\tilde{\rho}_{\rm k}=\rho_{\rm k}/3H_0^2=\Omega_{\rm
k_0}a^{-2}$ and the continuity equation for the curvature
\begin{eqnarray}
\tilde{p}_{\rm k}=-\tilde{\rho}_{\rm k}-\frac{1}{3}\frac{d
\tilde{\rho}_{\rm k}}{d x},
\end{eqnarray}
we obtain the curvature pressure
\begin{eqnarray}
\tilde{p}_{\rm k}=-\frac{1}{3}\Omega_{\rm k_0}a^{-2}.
\end{eqnarray}
By considering
$\tilde{p}_{\Lambda_0}=\omega_0\tilde{\rho}_{\Lambda_0}=\omega_0\Omega_{\Lambda_0}$,
one can obtain the following relations between the three constants
$\alpha$, $\beta$ and $C$ appeared in Eqs. (\ref{rholambda2}) and
(\ref{plambda}) as
\begin{eqnarray}
C=1&+&\frac{2\Omega_{\rm
m_0}}{2(\Omega_{\Lambda_0}-1)+\beta[3\Omega_{\rm m_0}+4\Omega_{\rm
r_0}+3(1+\omega_0)\Omega_{\Lambda_0}+2\Omega_{\rm k_0}-3]}
\nonumber\\
&+&\frac{2\Omega_{\rm
r_0}}{2(\Omega_{\Lambda_0}-1)+\beta[3\Omega_{\rm m_0}+4\Omega_{\rm
r_0}+3(1+\omega_0)\Omega_{\Lambda_0}+2\Omega_{\rm k_0}-4]}
\nonumber\\
&+&\frac{2\Omega_{\rm
k_0}}{2(\Omega_{\Lambda_0}-1)+\beta[3\Omega_{\rm m_0}+4\Omega_{\rm
r_0}+3(1+\omega_0)\Omega_{\Lambda_0}+2\Omega_{\rm
k_0}-2]},\label{C}
\end{eqnarray}
and
\begin{equation}
\alpha=\frac{1}{2}\Big[2\Omega_{\Lambda_0}+\beta\Big(3\Omega_{\rm
m_0}+4\Omega_{\rm
r_0}+3(1+\omega_0)\Omega_{\Lambda_0}+2\Omega_{\rm
k_0}\Big)\Big].\label{alpha}
\end{equation}
The deceleration parameter is given by
\begin{equation}
q=\frac{1}{2}+\frac{3}{2}\Big(\frac{\tilde{p}_{\rm
r}+\tilde{p}_{\Lambda}+\tilde{p}_{\rm k}}{\tilde{\rho}_{\rm
m}+\tilde{\rho}_{\rm r}+\tilde{\rho}_{\Lambda}+\tilde{\rho}_{\rm
k}}\Big),\label{q}
\end{equation}
where $p_{\rm m}=0$ for dust matter, $\tilde{\rho}_{\rm
m}=\rho_{\rm m}/3H_0^2=\Omega_{\rm m_0}a^{-3}$, $\tilde{\rho}_{\rm
r}=\rho_{\rm r}/3H_0^2=\Omega_{\rm r_0}a^{-4}$, and
$\tilde{p}_{\rm r}=\frac{1}{3}\Omega_{\rm r_0}a^{-4}$. In what
follows we neglect the contribution from radiation, i.e.
$\Omega_{\rm r_0}=0$.

Note that if one set $\Omega_{\rm k_0}=0$, then Eqs. (\ref{F3}),
(\ref{rholambda2}), (\ref{plambda}) ,(\ref{C}) to (\ref{q}) reduce
to Eqs. (2.5), (2.6), (2.8) to (2.11) in \cite{Granda1},
respectively.
\section{Numerical results}
\begin{figure}
\includegraphics{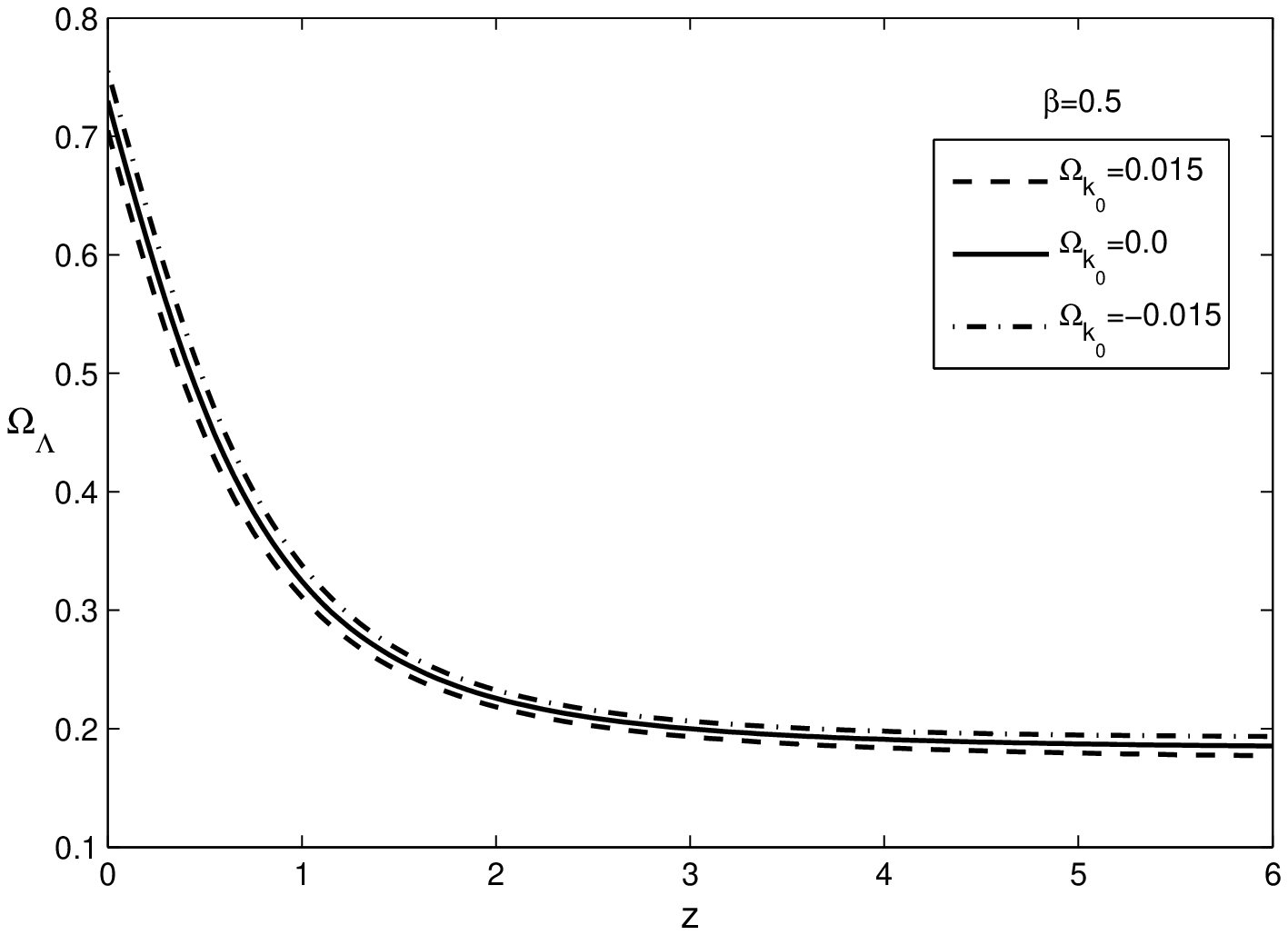}
      \vspace{6.5cm}
      \caption[]{DE density parameter versus redshift for open ($\Omega_{\rm k_0}=0.015$), flat ($\Omega_{\rm k_0}=0.0$)
      and closed ($\Omega_{\rm k_0}=-0.015$) universes.
      Auxiliary parameters are: $w_{0}=-1$, $\Omega_{\rm m_0}=0.27$, $\Omega_{\rm r_0}=0$, and $\beta=0.5$.}
         \label{OmegaLambdaz-0-015}
   \end{figure}

\begin{figure}
\includegraphics{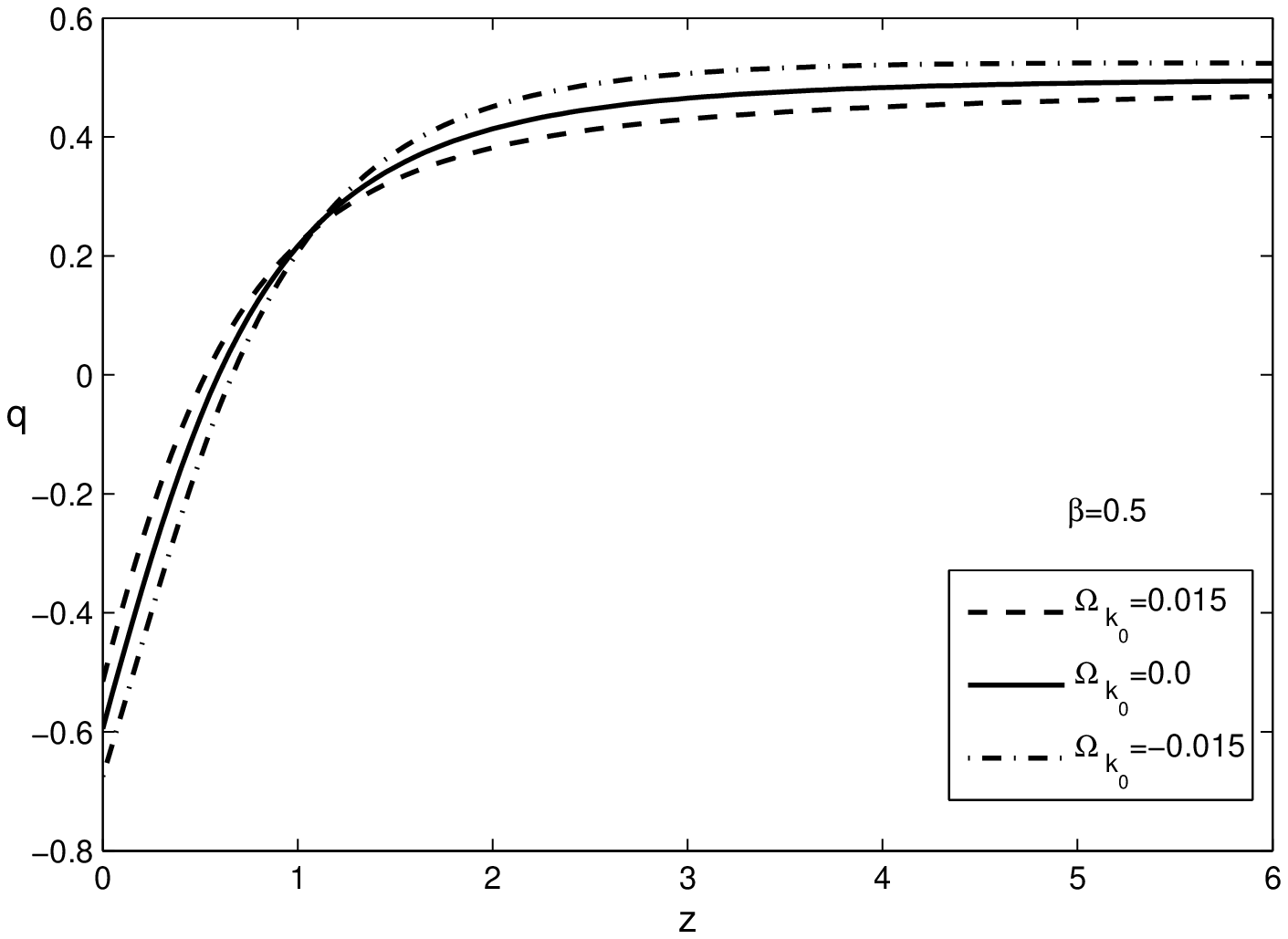}
      \vspace{6.5cm}
      \caption[]{Deceleration parameter versus redshift for open ($\Omega_{\rm k_0}=0.015$), flat ($\Omega_{\rm k_0}=0.0$)
      and closed ($\Omega_{\rm k_0}=-0.015$) universes.
      Auxiliary parameters as in Fig. \ref{OmegaLambdaz-0-015}.}
         \label{qz-0-015}
   \end{figure}

\begin{figure}
\includegraphics{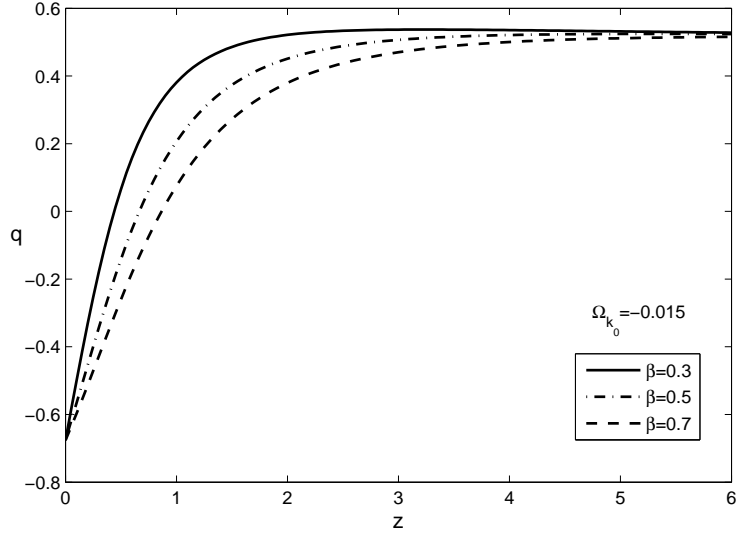}
      \vspace{6.5cm}
      \caption[]{Deceleration parameter versus redshift for closed universe with $\Omega_{\rm k_0}=-0.015$ and for the
      different values of $\beta$=0.3 (solid line), 0.5 (dash-dotted line) and 0.7 (dashed line).
      Auxiliary parameters are: $w_{0}=-1$, $\Omega_{\rm m_0}=0.27$, and $\Omega_{\rm r_0}=0$.}
         \label{qz-3beta-015}
   \end{figure}

\begin{figure}
\includegraphics{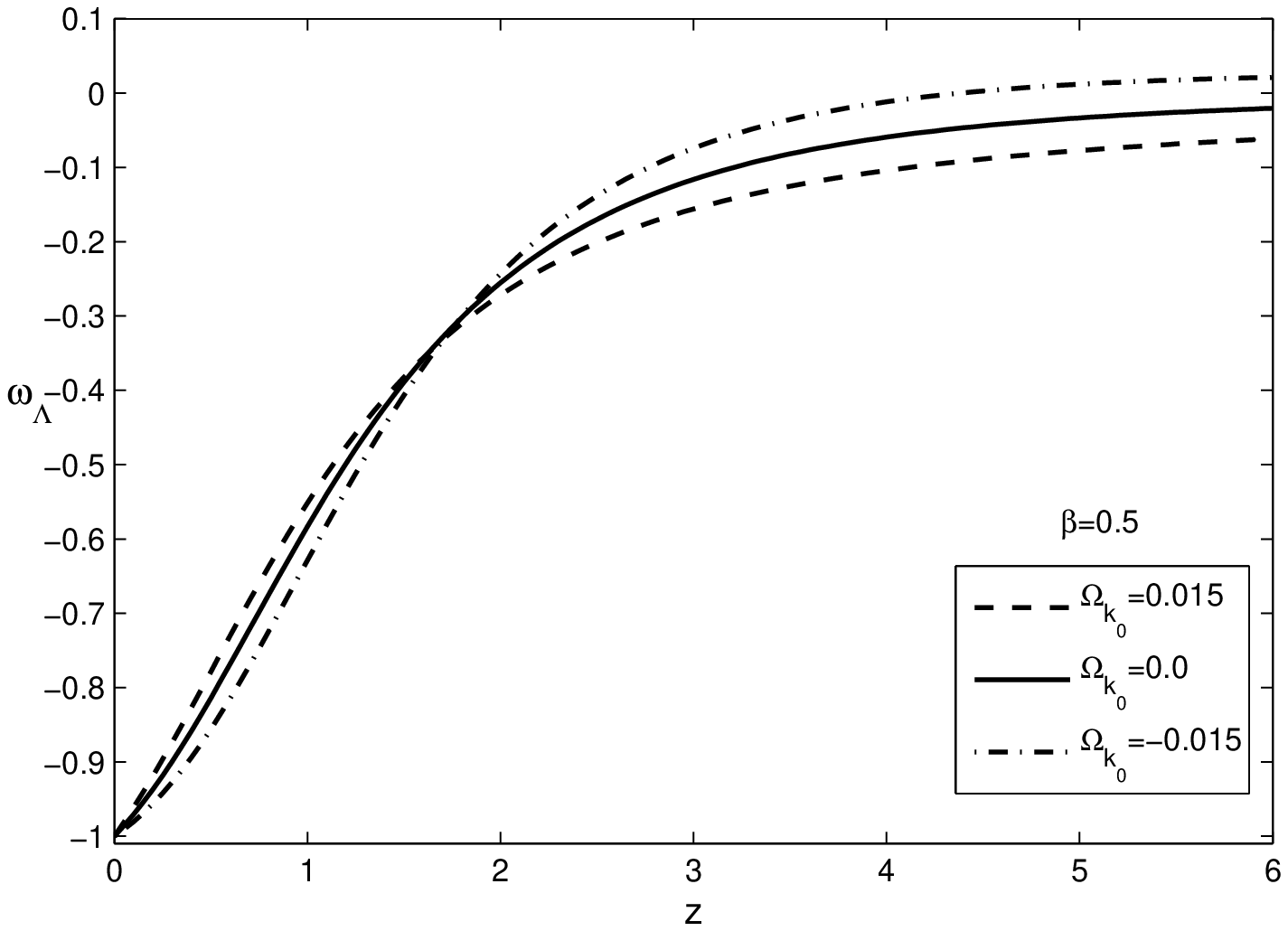}
      \vspace{6.5cm}
      \caption[]{EoS parameter versus redshift for open ($\Omega_{\rm k_0}=0.015$), flat ($\Omega_{\rm k_0}=0.0$)
      and closed ($\Omega_{\rm k_0}=-0.015$) universes.
      Auxiliary parameters as in Fig. \ref{OmegaLambdaz-0-015}.}
         \label{w-0-015}
   \end{figure}

\begin{figure}
\includegraphics{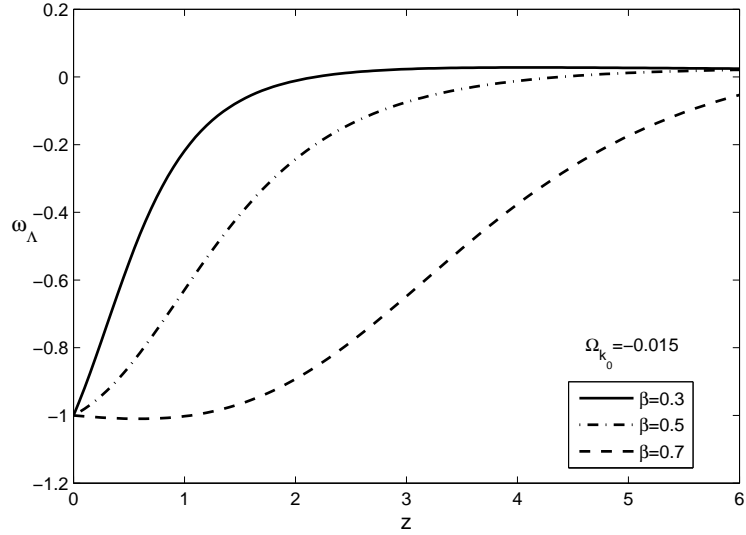}
      \vspace{6.5cm}
      \caption[]{EoS parameter versus redshift for closed universe with $\Omega_{\rm k_0}=-0.015$ and for the
      different values of $\beta$=0.3 (solid line), 0.5 (dash-dotted line) and 0.7 (dashed line).
      Auxiliary parameters as in Fig. \ref{qz-3beta-015}.}
         \label{w-3beta-015}
   \end{figure}

The evolution of the DE density parameter $\Omega_{\Lambda}$ given
by Eq. (\ref{OmegaLambda}), the declaration parameter $q$ given by
Eq. (\ref{q}) and the EoS parameter
$\omega_{\Lambda}=\tilde{p}_{\Lambda}/\tilde{\rho}_{\Lambda}$,
using Eqs. (\ref{rholambda2}) and (\ref{plambda}), are displayed
in Figs \ref{OmegaLambdaz-0-015}-\ref{w-3beta-015}. Figures
\ref{OmegaLambdaz-0-015} to \ref{qz-0-015} present the evolution
of the DE density parameter $\Omega_{\Lambda}$ and the declaration
parameter $q$ versus redshift $z$ for open ($\Omega_{\rm
k_0}=0.015$), flat ($\Omega_{\rm k_0}=0.0$) and closed
($\Omega_{\rm k_0}=-0.015$) universes with $\beta=0.5$. Choosing a
non-flat universe with $\Omega_{\rm k_0}=-0.015$ is compatible
with the recent observations \cite{Bennett}. Our numerical results
in Figs. \ref{OmegaLambdaz-0-015} to \ref{qz-0-015} show that in
the presence of small curvature ($\Omega_{\rm k_0}=\pm 0.015$),
the average difference between the non-flat and the flat cases is
order of $10^{-2}$.

Figure \ref{qz-3beta-015} shows the evolution of the deceleration
parameter versus redshift for closed ($\Omega_{\rm k_0}=-0.015$)
universe with the three different values of $\beta$. Note that
Fig. \ref{qz-3beta-015} clears that for $\beta=0.5$ and 0.7, the
values of the transition redshift $z_{\rm T}$ same as the flat
case ($\Omega_{\rm k_0}=0.0$) in \cite{Granda1}, are consistent
with the current observational data.

Figure \ref{w-0-015} presents the evolution of the EoS parameter
$\omega_{\Lambda}$ versus redshift $z$ for open ($\Omega_{\rm
k_0}=0.015$), flat
 ($\Omega_{\rm k_0}=0.0$) and
closed ($\Omega_{\rm k_0}=-0.015$) universes with $\beta=0.5$.
Figure \ref{w-0-015} same as Fig. \ref{qz-0-015} shows an average
difference $O(10^{-2})$ between the non-flat and the flat cases.
Figure \ref{w-0-015} shows that for the closed universe, the EoS
parameter $\omega_{\Lambda}$ from nearly $0$ at $z\simeq4.4$ to
$-1$ at $z\rightarrow 0$, same as the flat case in \cite{Granda1}
and open universe with $\Omega_{\rm k_0}=0.015$, behaves like some
scalar filed models of DE. Figure \ref{w-3beta-015} shows the
evolution of the EoS parameter versus redshift for closed
($\Omega_{\rm k_0}=-0.015$) universe with the three different
values of $\beta$.

\section{Conclusions}
We used a holographic DE model with new infrared cut-off proposed
by GO \cite{Granda1}, which includes a term proportional to
$\dot{H}$. Contrary to the holographic DE based on the event
horizon, this model depends on local quantities, avoiding in this
way the causality problem, and solved the coincidence problem.
Hence the proposed new infrared cut-off can be considered as a
viable phenomenological model of holographic density. We extended
GO model \cite{Granda1} to the non-flat case. However, some
experimental data have implied that our universe is not a
perfectly flat universe and that it possesses a small curvature
($\Omega_{\rm k_0}\sim -0.015$) \cite{Bennett}. Although it is
believed that our universe is flat, a contribution to the
Friedmann equation from spatial curvature is still possible if the
number of e-foldings is not very large \cite{Huang}. We obtained
the evolution of the DE density parameter $\Omega_{\Lambda}$, the
declaration parameter $q$ and the EoS parameter $\omega_{\Lambda}$
for a non-flat universe. In the limiting case of a flat universe,
the results are in exact agreement with those obtained by GO
\cite{Granda1}. Our numerical results show that

(i) for the non-flat universe with small curvature ($\Omega_{\rm
k_0}=\pm0.015$), the DE density, the declaration and the EoS
parameters show an average difference $O(10^{-2})$ in comparison
with the flat case;

(ii) for the closed universe with $\Omega_{\rm k_0}=-0.015$ and
for the declaration parameter with $\beta=0.5$ and 0.7, the values
of the transition redshift $z_{\rm T}$  same as the flat case are
consistent with the current observational data;

(iii) for the closed universe with $\Omega_{\rm k_0}=-0.015$, the
EoS parameter $\omega_{\Lambda}$ for $0\leq z<4.4$, same as the
flat and open ($\Omega_{\rm k_0}=0.015$) cases, behaves like some
scalar filed models of DE.
\\
\\
\noindent{{\bf Acknowledgements}}\\ The authors thank the
reviewers for very valuable comments. This work has been supported
financially by Research Institute for Astronomy $\&$ Astrophysics
of Maragha (RIAAM), Maragha, Iran.

\clearpage
\clearpage

\end{document}